\documentstyle{article}

\newcommand{\ket}[1]{{|#1\rangle}}

\author{Christof Zalka\footnote{Supported by
Schweizerischer Nationalfonds and LANL} \\ zalka@t6-serv.lanl.gov}

\title{Grover's quantum searching algorithm is optimal}

\begin{document}
\maketitle
\begin{abstract}
I show that for any number of oracle lookups up to about $\pi/4~
\sqrt{N}$, Grover's quantum searching algorithm gives the maximal
possible probability of finding the desired element. I explain why
this is also true for quantum algorithms which use measurements during
the computation. I also show that unfortunately quantum searching
cannot be parallelized better than by assigning different parts of the
search space to independent quantum computers.

\end{abstract}

\section{Quantum searching}
Imagine we have $N$ cases of which only one fulfills our
conditions. E.g. we have a function which gives 1 only for one out of $N$
possible input values and gives 0 otherwise. Often an analysis of the
algorithm for calculating the function will allow us to find quickly
the input value for which the output is 1. Here we consider the case
where we do not know better than to repeatedly calculate the function
without looking at the algorithm, e.g. because the function is
calculated in a black box subroutine into which we are not allowed to
look. In computer science this is called an oracle. Here I consider
only oracles which give 1 for exactly one input. Quantum searching for
the case with several inputs which give 1 and even with an unknown
number of such inputs is treated in \cite{boyer}.

Obviously on a classical computer we have to query the oracle on
average $N/2$ times before we find the answer. Grover \cite{grover}
has given a quantum algorithm which can solve the problem in about
$\pi/4~ \sqrt{N}$ steps. Bennett et al. \cite{bennett} have shown that
asymptotically no quantum algorithm can solve the problem in less than
a number of steps proportional to $\sqrt{N}$. Boyer et
al. \cite{boyer} have improved this result to show that e.g. for a
50\% success probability no quantum algorithm can do better than only
a few percent faster than Grover's algorithm. I improve the proof,
showing that for any number of oracle lookups Grover's algorithm is
exactly (and not only asymptotically) optimal.

The abovementioned proofs have shown that asymptotically proportional
to $\sqrt{N}$ steps are necessary for quantum searching. They have
not said whether these steps can only be carried out consecutively or
whether they could (partially) be done in parallel. If they could be
done in parallel then a quantum computer containing $S$ oracles (thus
$S$ physical black boxes) running for $T$ time steps could search a
search space of $O(S^2 T^2)$. Now any unstructured search problem can
simply be parallelized by assigning different parts of the search
space to independent search engines (whether quantum or
classical). But using this ``trivial'' parallelization we can only
search a search space of $O(S T^2)$ using $S$ independent quantum
computers running Grover's algorithm. I show that we can not do
better.


Grover's quantum searching algorithm distinguishes between the $N$ possible
one-yes oracles (yes = 1) using $\pi/4~ \sqrt{N}$ oracle calls (that is,
evaluations of the function). It makes the same sequence of operations 
$\pi/4~ \sqrt{N}$ times. The sequence consists of 4 simple operations. The
input state to the algorithm is the (easily constructed) uniform amplitude
state:

\begin{equation}
\mbox{initial state} \qquad 
\phi_0 = \frac{1}{\sqrt{N}} \sum_{x=0}^{N-1} \ket{x}
\end{equation}
The $\ket{x}$ are the $N=2^l$ computational basis states (where every one of
the $l$ qubits is either 0 or 1) which correspond to the possible inputs to
the oracle. Thus for Grover's algorithm $N$ has to be a power of 2. Again
\cite{boyer} have generalized this to arbitrary $N$.

The 4 operations of Grover's algorithm are then:

\begin{eqnarray*}
1. && \qquad \ket{y} \to -\ket{y} \qquad \mbox{for the one marked $y$} \\
2. && \qquad H^l \\
3. && \qquad \ket{x} \to -\ket{x} \qquad \mbox{for all} \qquad x \not = 0 \\
4. && \qquad H^l
\end{eqnarray*}
The first step is really the invocation of the oracle. The input we
are looking for is $y$. An oracle giving 0 or 1 can easily be changed
into an oracle which conditionally changes the sign of the input. This
can be done by preparing the qubit into which the oracle output bit
will be XORed in the state $(\ket{0}-\ket{1})/\sqrt{2}$.  Of course
the oracle will need work space, but as we expect these work
qubits to be reset to their pre-call value after each oracle call, we
do not really have to care about them in the quantum algorithm as they
``factor out'' (after the oracle call they form a tensor product with
the rest of the quantum computer). The requirement that the work
qubits have to be ``uncomputed'' means that the algorithm in the
oracle may take longer than its conventional irreversible version.

The second step applies a Hadamard transform to every one of the $l$
qubits (shorthand $H^l$). The Hadamard transform is given by the
following matrix:

\begin{equation}
H = \frac{1}{\sqrt{2}} 
\left(\begin{array}{rr} 1 & 1 \\ 1 & -1 \end{array} \right)
\end{equation}

The third step changes the sign of all computational basis states except for
$\ket{0}$ and the forth step is the same as the second.

The following results are straightforward to obtain: After any number
$i$ of iterations of these 4 operations the state of the QC can be
written as a linear combination of two fixed states:

\begin{equation}
\phi_i = A_i~ \frac{1}{\sqrt{N-1}} \sum_{x \not = y} \ket{x}
+ B_i~ \ket{y}
\end{equation}
One application of the 4 operations gives:

\begin{eqnarray*} 
A_{i+1} &=& (1-\frac{2}{N})~ A_i - 2 \frac{\sqrt{N-1}}{N}~ B_i \\
B_{i+1} &=& 2 \frac{\sqrt{N-1}}{N}~ A_i + (1-\frac{2}{N})~ B_i
\end{eqnarray*}
This is simply a SO(2) rotation with 

\begin{equation} \label{angle}
\cos(\varphi)=1-\frac{2}{N} \qquad \mbox{and} \qquad 
\sin(\varphi)= 2 \frac{\sqrt{N-1}}{N}
\end{equation}
Thus $\varphi \approx \sin{\varphi} \approx 2/\sqrt{N}$ and therefore
after $\pi/4~ \sqrt{N}$ steps we obtain a state very close to
$\ket{y}$.

It turns out (and is easy to check) that the initial (uniform
amplitude) state can be written in terms of half the above angle:

\begin{equation}
\phi_0 = \cos(\varphi/2)~ \frac{1}{\sqrt{N-1}} \sum_{x \not = y} \ket{x}
+ \sin(\varphi/2)~ \ket{y}
\end{equation}
Thus the success probability after $T$ oracle calls is exactly 

\begin{equation} \label{prob}
p_T = \sin^2(T \varphi+\varphi/2) \qquad .
\end{equation}
(Note that after about $\pi/4~ \sqrt{N}$ iterations of Grover's
algorithm the success probability goes down again!)

A noteworthy remark: Actually because towards the end the success
probability goes very slowly to 1, if we want to minimize the
\underline{average} number of steps\footnote{as opposed to the maximum
(worst case) number of steps}, it pays off to end the computation
earlier and run the risk to have to start over. A simple calculation
shows that the
 average number can thereby be reduced by some 12.14\%
relative to the above result.

\section{tight bound on quantum searching}
Here I sketch my version of the proof from \cite{boyer} which gave the
tightest limit on quantum searching so far. It is an extension of the
earlier $\Omega(\sqrt{N})$ proof in \cite{bennett}. Later I obtain my
results by further refining the same proof.

In the proof I assume a quantum computation consisting only of unitary
transformations (plus the final measurement) without measurements
during the computation. This can be done without loss of generality:
Clearly a measurement of a qubit whose outcome will not be used to make
decisions on what further unitary transformations should be applied,
can be delayed to the end. For a practical QC it seems likely that
what further unitary transformations will be applied will depend on
outcomes of intermediary measurements, like e.g. in error
correction. Thus we will probably have a ``hybrid'' quantum-classical
computer, where the classical part reads measurement outcomes and
depending on that, controls the exterior fields that induce unitary
transforms on the qubits. The point now is that \underline{in
principle} the classical part can simply be replaced by quantum
hardware which does the same. This may use more space as we are now
restricted to reversible computation, but for what concerns us, it of
course does not increase the number of oracle invocations, and this is
really all we care about here.

So we do not care about the cost of any other unitary transforms,
actually we do not even ask whether they can efficiently be composed
of elementary gates. Of course once we have established that even
under this general viewpoint Grover's algorithm is optimal, we know
that the ``auxiliary'' unitary transforms can be realized with just a
few elementary gates. Actually I expect that in any sensible
application of Grover's algorithm these auxiliary operations are going
to be much easier (and faster) than the oracle call.

The proof gives a limit on the success probability achievable with $T$
(for time) oracle lookups. Thereby we average over the $N$ possible
oracles. In computer science one is usually interested in the worst
case, that is the oracle for which the success probability is the
smallest. Because the worst case probability is smaller or equal to
the average case probability we also get an upper limit on the
former. In Grover's algorithm the success probability is independent
of the oracle so the worst case and average case probabilities are the
same and thus Grover's algorithm is also optimal for the worst case
probability.

The proof works by analyzing how the difference of the QC states
between the cases when we have a specific one-yes oracle and when we
have the empty oracle (always giving 0) evolves. For the empty oracle
case I denote the QC state after $i$ oracle invocations by $\phi_i$
whereas $\phi^y_i$ denotes this state when we have an oracle that
gives yes only for input $y$. More precisely, these are the QC states
just before the next oracle call, thus in one register of the QC there
must be the input to the oracle.

The proof consists of two parts. The central part of the proof gives a
bound on how far from the empty-oracle case the state can have
diverged after $T$ oracle calls when the oracle has one yes. To get a
meaningful statement we have to average over all possible one-yes
oracles, as for any given $y$ a special algorithm could be made that
would do especially well for this case. The statement is:

\begin{equation} \label{central}
\sum_{y=0}^{N-1} |\phi^y_T-\phi_T|^2 \le 4 T^2
\end{equation}
The second part of the proof gives an upper bound on the success
probability $p$ in terms of the left hand side of the above equation:

\begin{equation} \label{ieq}
2 N-2 \sqrt{N} \sqrt{p} - 2 \sqrt{N}\sqrt{N-1} \sqrt{1-p}~~
 \stackrel{1.}{\le}~~ \sum_{y=0}^{N-1} |\phi^y_T-\phi_T|^2
\end{equation}
Where the ``1.'' above the ``$\le$'' is for later discussion. I prove
this inequality in the appendix, as it is not central to the
understanding of the proof.

Both inequalities together then give the desired lower bound on $T$ in
terms of $N$ and $p$. Asymptotically and for $p=1$ the statement is $T
\geq \sqrt{N/2}~$. I now derive the ``central'' inequality
\ref{central}. To simplify the notation I assume that, like in
Grover's algorithm, in every iteration (each containing one oracle
invocation) the quantum computer makes the same sequence of
operations. It is easy to see that the proof works just as well
without this restriction (just add additional indices to $U$ and
$U_y$). By

\begin{equation}
\Delta U = U - U_y
\end{equation}
I denote the difference between the unitary transformation
corresponding to the empty oracle and the unitary transformation
corresponding to the oracle giving 1 only for input $y$. The
transformations $U$ and $U_y$ will act identically on all
computational basis states except those where the register holding the
input to the oracle is in state $y$. To get an upper bound on
$|\phi^y_T-\phi_T|$, consider the following:

\begin{eqnarray} \label{cent}
\phi_T &=& (U_y+\Delta U) \phi_{T-1} = 
U_y(U_y+\Delta U) \phi_{T-2}+\Delta U \phi_{T-1} = \dots \\
&=& \phi^y_T + \sum_{i=0}^{T-1} (U_y)^i \Delta U \phi_{T-1-i}
\end{eqnarray}
Then

\begin{eqnarray*}
|\phi^y_T-\phi_T| &=& \left|\sum_{i=0}^{T-1} (U_y)^{T-1-i} 
\Delta U \phi_i \right|~ 
\stackrel{2.}{\le}~ \sum_i~ | (U_y)^{T-1-i} \Delta U \phi_i| =\\ 
&& \qquad = 
\sum_i |\Delta U \phi_i|~ \stackrel{3.}{\le}~ 2 \sum_i |P_y \phi_i|
\end{eqnarray*}
Where $P_y$ is the projector onto those computational basis states
which are going to query the oracle on input $y$. The numbering of the
inequality signs is again for later discussion. For the next step I
need the inequality $(\sum a_i)^2 \le T \sum a_i^2$, where the $a_i$'s
are any T real numbers. It follows from the equality

\begin{equation}
\left( \sum a_i \right) ^2 +\frac{1}{2}~ \sum_{i,j} (a_i-a_j)^2 = T \sum a_i^2
\end{equation}
which is easy to verify. So now we get:

\begin{equation} \label{before_sum}
|\phi^y_T-\phi_T|^2 \le \left( 2 \sum_i~ |P_y \phi_i| \right) ^2~
\stackrel{4.}{\le}~ 4 T \sum_i~ |P_y \phi_i|^2
\end{equation}
By summing this over all $y$'s we get:

\begin{equation} \label{bound}
\sum_{y=0}^{N-1} |\phi^y_T-\phi_T|^2 \le 
 4 T \sum_i \underbrace{\sum_y |P_y \phi_i|^2}_{=1} = 4 T^2
\end{equation}

\section{improving this bound}

To see how tight the above inequality is, let us look at the 4
(numbered) inequalities in the proof which are then concatenated to
yield the final inequality. Let us see how well Grover's algorithm
(after any number of steps and for any marked $y$) does on these 4
inequalities. It turns out that it saturates all but the
2. inequality. For the 1. inequality this is shown in the
appendix. The 3. (in)equality is easily verified. It is true because
in the first of Grover's 4 operations the sign of $\ket{y}$ is
changed, thus maximizing the distance between $\ket{y}$ and
$-\ket{y}$. The 4. inequality is saturated because for Grover's
algorithm the $\phi_i$'s are all identical. So let us now concentrate
on the 2. inequality:

\begin{equation}
|\phi^y_T-\phi_T| = \left|\sum_i~ (U_y)^{T-1-i} \Delta U \phi_i \right|~ 
\stackrel{2.}{\le}~ \sum_i~ | (U_y)^{T-1-i} \Delta U \phi_i|
\end{equation}
For Grover's algorithm, $U$ is the identity and thus $\phi_i=\phi_0$, so 

\begin{equation}
\Delta U~ \phi_i = 2 P_y~ \phi_i = \frac{2}{\sqrt{N}}~ \ket{y}  
\end{equation}
As mentioned before, $U_y$ just carries out a SO(2) rotation on the
space spanned by $\ket{y}$ and $1/\sqrt{N-1}~~ \sum_{x \not= y}
\ket{x}$. Thus for Grover's algorithm the vectors in inequality \#2
do not all point in the same direction, rather if drawn one after the
other (to form the vector sum) they form an arc. This prevents the
inequality from being saturated and explains the discrepancy of the
tight bound in \cite{boyer} from the performance of Grover's
algorithm.

So let us try to find a tighter (upper) bound on
$|\phi^y_T-\phi_T|$. To this end I write equation \ref{cent} a little
bit differently (where of course $T$ does not mean ``transpose''!):

\begin{eqnarray*}
\phi_T=(U~ \phi_{T-1}-U_y~ \phi_{T-1})&+&
(U_y U~ \phi_{T-2}-U_y U_y~ \phi_{T-2})+ \dots \\
&&\dots + ((U_y)^{T-1} U~ \phi_0 - (U_y)^T~ \phi_0) ~+~ \phi^y_T
\end{eqnarray*}
This has the form 

\begin{equation}
\psi_0 - \psi_T=(\psi_0-\psi_1)+(\psi_1-\psi_2)+(\psi_2-\psi_3)+\dots 
+(\psi_{T-1}-\psi_T)
\end{equation}
where all $\psi_i$'s are normalized. The question now is how we have
to choose $\psi_1, \psi_2, \dots \psi_{T-1}$ in order to minimize
$\sum|\psi_i-\psi_{i+1}|^2$ when $\psi_0$ and $\psi_T$ are fixed. Note
that the relative phases of different states of the QC have no
physical meaning and are therefore only a matter of convention. This
is because quantum states are really given by 1-dimensional subspaces
(rays) of a Hilbert space and not by vectors. We can thus assume that
$\langle \psi_0 \ket{\psi_T}$ is real and
non-negative.\footnote{Actually it would be nicer to write the proof
in terms of absolute values of scalar products only, thus avoiding
unphysical quantities like the difference of state vectors}
Intuitively it is clear that for the minimum the $\psi_i$'s have to be
evenly spaced along the arc between $\psi_0$ and $\psi_T$. (Note that
this is obviously the case for Grover's algorithm!) Formally this can
be established by setting the derivative with respect to the
components of $\psi_i$ equal to zero, which yields:

\begin{equation} \label{arc}
\psi_i=\frac{\psi_{i-1}+\psi_{i+1}}{|\psi_{i-1}+\psi_{i+1}|}
\end{equation}

We can imagine all these vectors to lie in a 2-dimensional real vector
space spanned by $\psi_0$ and $\psi_T$. From planar trigonometry we
then get (draw a picture with a line bisecting the angle between
$\psi_0$ and $\psi_T$):

\begin{equation} \label{arc2}
|\psi_0-\psi_T|^2=\left( 2 \sin (\frac{\alpha}{2}) \right)^2 \quad 
\Rightarrow \quad
|\psi_i-\psi_{i+1}|^2=\left( 2 \sin (\frac{\alpha}{2 T}) \right)^2
\end{equation}
Where $\alpha$ is the angle between $\psi_0$ and $\psi_T$ and
$\alpha/T$ the angle between $\psi_i$ and $\psi_{i+1}$ for all $i$.

So we now have that (compare to equation \ref{before_sum})

\begin{equation} \label{before_sum2}
|\phi_T-\phi_T^y|^2~ \le~ f \left(~ 4~ T \sum_i |P_y \phi_i|^2~ \right) 
\qquad \forall~ y
\end{equation}
where $f(x)$ describes the improvement in our bound. It is given by
($T$ is a fixed parameter)

\begin{equation}
f (~ \underbrace{4~ T^2 \sin^2 (\frac{\alpha}{2 T})} _{\approx~
\alpha^2} ~ ) = \underbrace{4 \sin^2 (\alpha/2)}_{\le~ \alpha^2}
\end{equation}

We now want to sum equation \ref{before_sum2} over all $y$'s. We use
calculus to get an upper bound on the sum over the right hand side. I
claim:

\begin{equation} \label{imp}
\sum_y |\phi_T-\phi_T^y|^2~ 
\le~ \sum_y f \left(~ 4~ T \sum_i |P_y \phi_i|^2~ \right) 
\le~ N~ f \left(~ \frac{1}{N} \sum_y~ 4~ T \sum_i |P_y \phi_i|^2~ \right) 
\end{equation}
We know that we have an absolute (and not only a relative) maximum on
the right hand side because in the whole area of interest $f'>0$ and
$f''<0$. More precisely, $f'>0$ is true exactly as long as the number
of steps $T$ in Grover's algorithm is below the (fractional) optimum
(which can be obtained from equation \ref{prob}) of about $\pi/4
\sqrt{N}$. The above inequality is saturated by Grover's algorithm as
there the above optimal situation with equal angles between successive
vectors is realized with this constant angle equal to $\varphi$ given
in equation \ref{angle}. Thus we have established that Grover's
algorithm is optimal.

\section{Limits on parallelizing Grover's algorithm}

Assume we have $S$ ($S$ for ``space'') identical oracles with exactly
one marked element. Thus we can imagine that we have $S$ identically
constructed physical black boxes. In particular I assume that all
oracles take the same time to answer a query.

We want to find the fastest way to obtain the marked element with a
quantum computer that is allowed to use all these oracles (and may
input entangled states to them). To formalize this in a ``query
complexity'' way and assuming that querying will take much more time
than the other operations in the algorithm, I only consider the
``querying time'', which is the time during the algorithm when any one
of the oracles is working.

The quantum computer could query the individual oracles at any time,
in particular it can start querying an oracle while another one is
still running. In the following I give an argument that without loss
of potential power of the algorithm, we can assume that the oracles
are always queried synchronously.

First imagine that we have just 2 oracles. We begin by querying the
first one and while it is still working we start querying the second
one. We can assume that while an oracle is working, only the oracle
interacts with its input register. (If necessary, this can be assured
by XORing the input state to a register reserved for the oracle.) Then
it follows that the second oracle could be queried as soon as the
first one, possibly by doing some preparatory gates ahead of time.

Now imagine we have $S$ oracles. First consider the very first
querying of an oracle in the algorithm. All the oracles which we start
querying while the first one is still working can, by the above
argument, be queried simultaneously with the first one. As there is no
point in not using the other oracles during this time, we can assume
that we start the algorithm by querying all $S$ oracles
simultaneously.

By applying the same argument to what happens after this first step,
we get that we can assume that the second step also consists of
querying all $S$ oracles simultaneously. By iterating this we see that
we can assume that the algorithm always queries all oracles
simultaneously. Say it does this $T$ times.

As before (equation \ref{imp}) we have

\begin{equation}
\sum_y |\phi_T-\phi_T^y|^2~ 
\le~ N f\left( \frac{1}{N} 4 T \sum_y \sum_i |P_y \phi_i|^2 \right)
\le~ 4 T \sum_y \sum_i |P_y \phi_i|^2
\end{equation}
Where on the right I have also included the old unimproved
result. Here $\phi_i$ is the QC state just before the $S$ oracles are
called (for the $i+1$ -st time). Now $P_y$ is the projector onto those
computational basis states where any oracle (possibly several ones) is
queried on input $y$.

It is easy to see that:

\begin{equation} \label{proj}
|P_y \phi_i|^2 ~\leq~ \sum_{k=1}^S |P_y^k \phi_i|^2 
\end{equation}
Here $P_y^k$ is the projector onto those computational basis states
where oracle number $k$ is queried on input $y$. The inequality
becomes an equality when there are no basis states in $\phi_i$ where
several oracles are queried on input $y$.

From that we get (compare to equation \ref{bound}):

\begin{equation}
\sum_y \sum_{i=0}^{T-1} |P_y \phi_i|^2 ~\leq~ 
\sum_{k=1}^S \sum_{i=0}^{T-1} \underbrace{\sum_y |P_y^k \phi_i|^2}_
{= 1} = S \cdot T
\end{equation}
So we get the final result

\begin{equation}
\sum_y |\phi_T-\phi_T^y|^2~ 
\le~ N f\left( \frac{1}{N} 4 T^2 S \right)
\le~ S \cdot 4 T^2
\end{equation}
This shows that to get a certain success probability we can gain only
a factor of $\sqrt{S}$ in $T$ by using $S$ oracles, but this is
essentially the same performance as $S$ independent Grover searches,
each working on one $S^{th}$of the total search space. If $N$ (the
size of the search space) is divisible by $S$ this is an exact
statement, otherwise we still get an asymptotic statement. \\ \\

\section{Appendices}
\subsection{proof of inequality \ref{ieq}}

The situation is as follows: a quantum system is in one of $N$ pure
states given by the normalized vectors $\psi_y$, $y=0\dots N-1$. The
task is to find in which of these states it is by using any
measurement procedure allowed by quantum theory. It is well known that
if the $\psi_y$'s are not all pairwise orthogonal, this can only be
done probabilistically (see e.g. \cite{fuchs}). Here we are interested
in maximizing the average success probability when averaging over all
$N$ cases. We want to prove the following upper bound on this
probability $p$:

\begin{equation} \label{ieq2}
2 N-2 \sqrt{N} \sqrt{p} - 2 \sqrt{N}\sqrt{N-1} \sqrt{1-p}~~
 \le~~ \sum_{y=0}^{N-1} |\psi_y-\psi|^2
\end{equation}
Measurement schemes can in general be such that when the measurement
gives $y$ then one is sure that the state was $\psi_y$, but then in
general the answer can also be ``I do not know''. Here we are not
interested in such schemes. Because once we get an answer we can
easily check whether it is correct, all we are interested in, is to
maximize the probability of getting the right answer, irrespective of
whether an unsuccessful measurement yields a wrong answer or ``do not
know''.

\subsubsection{Grover's algorithm saturates the inequality}

Here the $\psi_y$'s are the different final states (depending on the
oracle) of the QC just before measurement and $\psi$ is the state we
get for the ``zero''-oracle. After any number of iterations of
Grover's algorithm these states can be written in terms of some $p$
as:

\begin{equation}
\psi_y=\sqrt{p}~ \ket{y} 
+\sqrt{1-p}~ \frac{1}{\sqrt{N-1}} \sum_{y' \not= y} \ket{y'} 
\qquad \mbox{and} \qquad \psi=\frac{1}{\sqrt{N}} \sum_y \ket{y}
\end{equation}
It is now easy to verify that inequality \ref{ieq2} is saturated.

\subsubsection{proof of the inequality}

In general the Hilbert space of the QC will have dimension $M > N$. We
must assume this because it may in general gives the possibility of a
measurement with a larger success probability. On the other hand a von
Neumann measurement on such an enlarged space {\it is} really the best
we can do (\cite{fuchs}). A von Neumann measurement is just a standard
quantum measurement given by a hermitian operator or, essentially
equivalently, by a set of (mutually orthogonal) eigenspaces which
together span the whole Hilbert space. I write the $\psi_y$'s and
$\psi$ as follows in terms of some basis $\ket{m}$:

\begin{equation}
\psi_y= \sum_{m=0}^{M-1} c^y_m \ket{m} \qquad \mbox{and} \quad
\psi= \sum_{m=0}^{M-1} c_m \ket{m}
\end{equation}
Without loss of generality we can assume that we measure in the basis
$\ket{m}$. I denote by $M_y$ the set of $m$'s which, when we obtain
them from a measurement, will be interpreted as answer ``$y$''.  By
$p_y$ I denote the probability of therby correctly identifying the state
$\psi_y$:

\begin{equation}
p_y=\sum_{m \in M_y} |c^y_m|^2
\end{equation}
To prove inequality \ref{ieq2} we look for the minimal value its right
hand side can assume for a given $p=1/N \sum_y p_y$. We do the
minimization in two steps, First we find the $\psi_y$ (= the
$c^y_m$'s) for which $|\psi_y-\psi|^2$ is minimal for a given fixed
$p_y$ and $\psi$. Using the Lagrange multiplier technique to find
(tentative) extrema under some constraint we get the expression:

\begin{equation}
|\psi_y-\psi|^2 -\lambda_1 |\psi_y|^2 - \lambda_2 p_y = 
\sum_{m=0}^{M-1} |c^y_m -c_m|^2 -\lambda_1 \sum_{m=0}^{M-1} |c^y_m|^2 
 -\lambda_2 \sum_{m \in M_y} |c^y_m|^2
\end{equation}
where the first constraint (Lagrange multiplier $\lambda_1$) comes
from the requirement that $\psi_y$ be normalized and the second
(Lagrange multiplier $\lambda_2$) because we want to minimize for
fixed $p_y$. To find candidate extrema we set the derivatives of this
expression with respect to the real and imaginary parts of $c^y_m$
equal to zero. The well known ``trick'' that in this case one can
formally treat the complex variable and its complex conjugate as the 2
independent real variables simplifies the calculation to obtain:

\begin{equation}
c_m = (1-\lambda_1-\lambda_2) c^y_m~~ \forall~ m \in M_y \qquad 
\mbox{and} \quad c_m = (1-\lambda_1) c^y_m~~ \forall~ m \not \in M_y
\end{equation}
By satisfying the constraints we get:

\begin{equation}
1-\lambda_1-\lambda_2= \pm \frac{\sqrt{a_y}}{\sqrt{p_y}} \qquad 
1-\lambda_1= \pm \frac{\sqrt{1-a_y}}{\sqrt{1-p_y}} \qquad 
\mbox{where} \quad a_y=\sum_{m \in M_y} |c_m|^2
\end{equation}
From that we get the following 4 candidates for a minimum:

\begin{equation}
|\psi_y-\psi|^2 = 2 -2\left(\pm \sqrt{p_y} \sqrt{a_y} 
\pm \sqrt{1-p_y} \sqrt{1-a_y}
\right)
\end{equation}
Of course the minimum is reached when both signs are positive. Note
that we do not have to worry that $|\psi_y-\psi|^2$ might be even
smaller on some boundary of the parameter range we minimized over.
This is because our parameter range (under the constraints) does not
have a boundary, thus it is a bona fide manifold. Also our coordinate
system is regular all over the manifold. By summing over $y$ we get:

\begin{equation} \label{middle}
\sum_y |\psi_y-\psi|^2 ~\geq~ 2 N -2 \sum_y 
\left( \sqrt{p_y} \sqrt{a_y} +\sqrt{1-p_y} \sqrt{1-a_y} \right)
\end{equation}
Now we look for the $\psi$ (= the $a_y$'s) and the $p_y$'s for which
this becomes minimal. We have the constraints $|\psi|^2=\sum |c_m|^2
=\sum a_y =1$ and $1/N \sum_y p_y =p$ for a fixed $p$.  Again using
the Lagrange multiplier technique we get $a_y=1/N~ \forall y$ and
$p_y=p~ \forall y$. Then:

\begin{equation}
\sum_{y=0}^{N-1} |\psi_y-\psi|^2 
= 2 N-2 \sqrt{N} \sqrt{p} - 2 \sqrt{N}\sqrt{N-1} \sqrt{1-p}~~
\end{equation}
This time the parameter range over which we have minimized does have a
boundary. The boundary is reached when one of the $a_y$'s or $p_y$'s
is either 0 or 1. Still we can show that we have really found a global
minimum by showing that the second derivative is positive definite
over the whole parameter range. To avoid having to adapt this argument
to the situation with constraints, we can e.g. set $p_0=N p-\sum_{y
\not = 0} p_y$ and $a_0=1-\sum_{y \not =0} a_y$ and then check that the
second derivative of the right hand side of equation \ref{middle} is
always positive definite.

\section{final remarks}

I have here only considered oracles with the promise that there is
exactly one marked element. It seems very plausible that the proof can
be extended to oracles with any known(!) number of marked
elements. The same may be true for the case where we have a
non-uniform a priori probability for the different 1-yes oracles and
we want to maximize the average success probability. Then one also has
to consider a modified Grover algorithm.

Also it seems that by reading the proof carefully, one can establish
that Grover's algorithm is essentially the only optimal algorithm.

As for any no-go theorem which claims implications for the physical
world, we must be careful about the assumptions we made. Arguably the
main assumption made here is that the time evolution of quantum states
is exactly linear as of course it is in standard quantum theory. Most
physicists think this is very likely.

I would like to thank Manny Knill and Daniel Gottesman for helpful
discussions.

\end{document}